\documentclass[twocolumn]{aastex6}
\RequirePackage{lineno}
\usepackage{amsmath}
\usepackage{amssymb}
\usepackage{amsthm}
\usepackage{natbib,aasdefs,url,bm}
\usepackage{array}
\usepackage{float}
\usepackage{graphicx}
\usepackage{color}

\newcounter{ichi}
\newcounter{ni}
\newcounter{san}
\newcounter{yon}
\newcommand{\grb}{GRB 160821B }
\newcommand{\vhe}{VHE $\gamma$-rays }
\newcommand{\RN}[1]{%
  \textup{\uppercase\expandafter{\romannumeral#1}}%
}

\definecolor{darkred}{rgb}{0.55, 0.0, 0.0}

\slugcomment{}

\shorttitle{High-energy gamma-ray emission from GRB 160821B}
\shortauthors{Zhang, Murase, Yuan, Kimura, M\'esz\'aros, and Fang}

\begin{document}

\title{External Inverse-Compton Emission Associated with Extended and Plateau Emission of Short Gamma-Ray Bursts: Application to GRB 160821B}
       
\author{
B. Theodore Zhang\altaffilmark{1,2,3}, 
Kohta Murase\altaffilmark{1,2,3,4}, 
Chengchao Yuan\altaffilmark{1,2,3},
Shigeo S. Kimura\altaffilmark{5,6},
Peter M\'esz\'aros\altaffilmark{1,2,3}
}
\altaffiltext{1}{Department of Physics, Pennsylvania State University, University Park, Pennsylvania 16802, USA}
\altaffiltext{2}{Department of Astronomy \& Astrophysics, Pennsylvania State University, University Park, Pennsylvania 16802, USA}
\altaffiltext{3}{Center for Multimessenger Astrophysics, Institute for Gravitation and the Cosmos, Pennsylvania State University, University Park, Pennsylvania 16802, USA}
\altaffiltext{4}{Center for Gravitational Physics, Yukawa Institute for Theoretical Physics, Kyoto University, Kyoto, Kyoto 606-8502, Japan}
\altaffiltext{5}{Frontier Research Institute for Interdisciplinary Sciences, Tohoku University, Sendai 980-8578, Japan}
\altaffiltext{6}{Astronomical Institute, Tohoku University, Sendai 980-8578, Japan}

\begin{abstract}
The recent detection of TeV photons from two gamma-ray bursts (GRBs), GRB 190114C and GRB 180720B, has opened a new window for multi-messenger and multi-wavelength astrophysics of high-energy transients. We study the origin of very-high-energy (VHE) $\gamma$-rays from the short GRB 160821B, for which the MAGIC Collaboration reported a $\sim 3 \sigma$ statistical significance. Short GRBs are often accompanied by extended and plateau emission, which is attributed to internal dissipation resulting from activities of a long-lasting central engine, and Murase et al. (2018) recently suggested the external inverse-Compton (EIC) scenario for VHE counterparts of short GRBs and neutron star mergers. Applying this scenario to GRB 160821B, we show that the EIC flux can reach $\sim 10^{-12}\rm~erg~cm^{-2}~s^{-1}$ within a time period of $\sim 10^3 - 10^4\rm~s$, which is consistent with the MAGIC observations. EIC $\gamma$-rays expected during the extended and plateau emission will be detectable with greater significance by future detectors such as the Cherenkov Telescope Array (CTA). The resulting light curve has a distinguishable feature, where the VHE emission is predicted to reach the peak around the end of the seed photon emission.
\end{abstract}

\keywords{non-thermal, gamma-ray bursts}


\section{\label{sec:intro}Introduction}
Very-high-energy (VHE) $\gamma$-rays (with energies higher than $\sim 0.1\rm~TeV$) represent the most energetic part of the currently observed EM spectrum. 
The \vhe play an important role in high-energy multi-messenger and multi-wavelength astrophysics~\citep[e.g.,][]{Inoue:2013vy,Murase:2019tjj, Meszaros:2019xej,Hinton:2019etp}.
Imaging Atmospheric Cherenkov Telescopes (IACTs) detect Cherenkov light that is produced during the development of extensive air showers as the \vhe enter the Earth's atmosphere~\citep[e.g.,][]{Hinton:2008ka, Lorenz:2012nw}.
The detection of \vhe from GRB 190114C by the Major Atmospheric Gamma Imaging Cherenkov (MAGIC) telescopes~\citep{Acciari:2019dxz, Acciari:2019dbx} and GRB 180720B by the High Energy Stereoscopic System (H.E.S.S.)~\citep{Arakawa:2019cfc} has opened a new window for the exploration of the physics of relativistic shocks involving particle acceleration~\citep{Meszaros:2006rc, Kumar:2014upa}. 

On the other hand, the discovery of the first double neutron star (NS) merger event GW170817, associated with GRB 170817A~\citep{Monitor:2017mdv}, is a milestone in the multi-messenger astronomy, which was initially detected through gravitational waves (GWs)~\citep{TheLIGOScientific:2017qsa}, and later observed through electromagnetic (EM) emission from radio to $\gamma$-rays~\citep{GBM:2017lvd}. 
VHE $\gamma$-ray emission has also been recently discussed in this context~\citep{Murase:2017snw,Kimura:2019fae}, and the GW follow-up observations by Cherenkov telescopes, e.g., the Cherenkov Telescope Array (CTA), is promising in the near future~\citep{Kakuwa:2011aq,Inoue:2013vy,Bartos:2019tsp}. 
In the synchrotron self-Compoton (SSC) scenario, the same population of electrons that emit synchrotron photons in the dissipation region of the relativistic outflow upscatter these photons to much higher energies by a factor of $\sim \gamma_e^2$, where $\gamma_e$ is the electron Lorentz factor~\citep[e.g.,][]{Meszaros:1994sd, Sari:2000zp,Wang:2001fu, Zhang:2001az, Acciari:2019dbx, Arakawa:2019cfc}. 
In the presence of long-lasting central engine activities, late-prompt photons that are related to the extended and/or plateau emission can be upscattered to the VHE band by high-energy electrons accelerated at the external forward shock via external inverse-Compton (EIC) emission~\citep[][for NS mergers]{Murase:2017snw}.
On the other hand, if the prolonged jets dissipate inside cocoon, \vhe via upscattering of thermal cocoon photons in the jet have also been predicted~\citep{Kimura:2019fae}.

The MAGIC experiment utilizes two IACTs to detect \vhe within the energy range from $\sim 50\rm~GeV$ to $\sim 50\rm~TeV$~\citep{Aleksic:2014poa, Aleksic:2014lkm}. Besides the confirmed detection of GRB 190114C, there are three short GRBs, GRB 061217, GRB 100816A, and GRB 160821B with redshifts $z < 1$ and time delays after the GRB trigger $< 1 \rm \ hour$ that have been followed-up by the MAGIC telescopes under adequate conditions~\citep{Acciari:2019dxz}.
In particular, the MAGIC Collaboration recently reported an excess of \vhe from the direction of the short \grb within the time window of 24 s to 4 hour after the trigger~\citep{Inoue:2019ICRC, Acciari:2020ljs}, and there was a $\sim 3\sigma$ signal at $t \sim 10^4\rm~s$. \grb is one of the nearest short GRBs identified by \textit{the Neil Gehrels Swift Observatory}~\citep{2016GCN.19846....1L}, which is located at the outskirts of the host galaxy with a measured redshift of $z \sim 0.16$ (or a luminosity distance of $\sim 780\rm~Mpc$). Although the detection is still tentative, \grb may be the first short GRB detected in the VHE band.

The $\gamma$-ray and X-ray light curves observed by \textit{Swift} often show a period of extended emission in the early X-ray afterglow phase with a duration of $\sim 300$ seconds~\citep{Norris:2006rw, Berger:2013jza}. 
The existence of the extended emission component supports the argument that the central engine activity lasts for a longer time than the prompt emission, which could be explained by a magnetar~\citep[][]{Dai:2006hj} or black hole accretion~\citep[][]{Kisaka:2015sya}.
Short GRBs with extended emission usually show a plateau emission phase at later times~\citep{Gompertz:2013aka,Gompertz:2013zga, Kisaka:2017tas}.
The X-ray flux during the plateau phase can be explained within the external forward shock model, considering a complicated structured jet with an appropriate initial Lorentz factor~\citep[][for GRB 160821B]{Troja:2019ccb,Lamb:2019lao}, or else through a refreshed shock scenario in which the forward shock is continuously replenished via the collision with the slower but more energetic portions of the ejecta~\citep{Matsumoto:2020fle}.
Alternatively, both extended and plateau emission components are attributed to late-prompt emission from the long-lasting internal dissipation~\citep{Ghisellini:2007bd,Murase:2010fq, Kisaka:2015sya,Kisaka:2017tas}, where the external shock emission can be out-shined by the late-prompt emission.

In this Letter, we explore the origin of \vhe from \grb considering both of the SSC and EIC scenarios. 
In Sec.~\ref{sec:SSC-scenario}, we show that the SSC scenario is disfavored to explain the observed \vhe given the multi-wavelength constraints. In Sec.~\ref{sec:EIC-scenario}, we present the results of the EIC scenario considering extended and plateau emission as seed photons.
We discuss the implications of our results and give a summary in Sec.~\ref{sec:summary}.

\section{VHE $\gamma$-rays in the synchrotron-self Compton scenario}\label{sec:SSC-scenario}
First, we consider the standard external forward shock model~\citep{Meszaros:1996sv,Sari:1997qe}, where electrons are accelerated to higher energies via the diffusive shock acceleration mechanism. 
In the SSC scenario, synchrotron photons emitted from these electrons are upscattered via the inverse-Compton (IC) emission mechanism. 
Following~\cite{Zhang:2020qbt}, the dynamics of the outflow is calculated by solving the 1D differential equation. 
The steady-state electron energy spectrum is derived by solving the continuity equation, and the observed flux from relativistic outflows is calculated via the integration along the equal-arrival-time surface (EATS). See~\cite{Zhang:2020qbt} for details.

The best-fit parameters utilizing the multi-wavelength data from radio to X-rays suggest relatively larger values of $\epsilon_B \sim 10^{-2}$~\citep{Lamb:2019lao,Troja:2019ccb}, where $\epsilon_B$ is the energy fraction of the internal energy that is converted into the magnetic energy. 
However, the resulting SSC flux turns out to be too low for $\epsilon_e/\epsilon_B \sim 10$
In this work, we adopt $\epsilon_e = 0.3$ and $\epsilon_B = 10^{-5}$ to achieve the SSC dominance ~\citep[e.g.,][]{Derishev:2019cgi,Arakawa:2019cfc}.
Note such a small value of $\epsilon_B$ can be consistent with the values extrapolated from the value obtained by numerical simulations~\citep{Vanthieghem:2020nvr}. 

The relation between the peak flux of IC and seed photons is approximately written as $F_{\varepsilon,\rm max}^{\rm IC} \sim \tau_e f_e F_{\varepsilon, \rm max}^{\rm seed}$ in the Thompson regime, where $\tau_e \sim \sigma_T n_{\rm ex} R$ is the Thomson optical depth in the shocked region, $\sigma_T$ is the Thomson cross section, $f_e$ is the number fraction of electrons that can be accelerated, $n_{\rm ex}$ is the number density of external material, and $R$ is the size of the emission region~\citep{Sari:2000zp,Wang:2001fu,Zhang:2001az}.
Thus, one expects that higher external density environments are preferred for strong IC emission.
Unlike long GRBs that may occur in dense environments as a result of extensive wind losses, short GRBs usually occur at the outskirts of the host galaxy due to the kick after the supernova explosions. 
Observations have revealed that the external matter density surrounding short GRBs usually have $n_{\rm ex} \sim 10^{-3} - 1 \rm~cm^{-3}$, depending on the merger timescale and kick velocity~\citep{Berger:2013jza}, and the mean value is $n_{\rm ex} \sim 3\times 10^{-3} - 1.5 \times 10^{-2} \rm~cm^{-3}$ ~\citep{Fong:2015oha}.
The best-fit parameters derived from the fitting to the multi-wavelength data of \grb in the forward shock model show that the external matter density can be as low as $\sim 10^{-4}\rm~cm^{-3}$~\citep{Lamb:2019lao, Troja:2019ccb}, but can still be as high as $n_{\rm ex} \sim 0.1\rm~cm^{-3}$~\citep{Schroeder:2020qbf}.

\begin{figure}
\includegraphics[width=\linewidth]{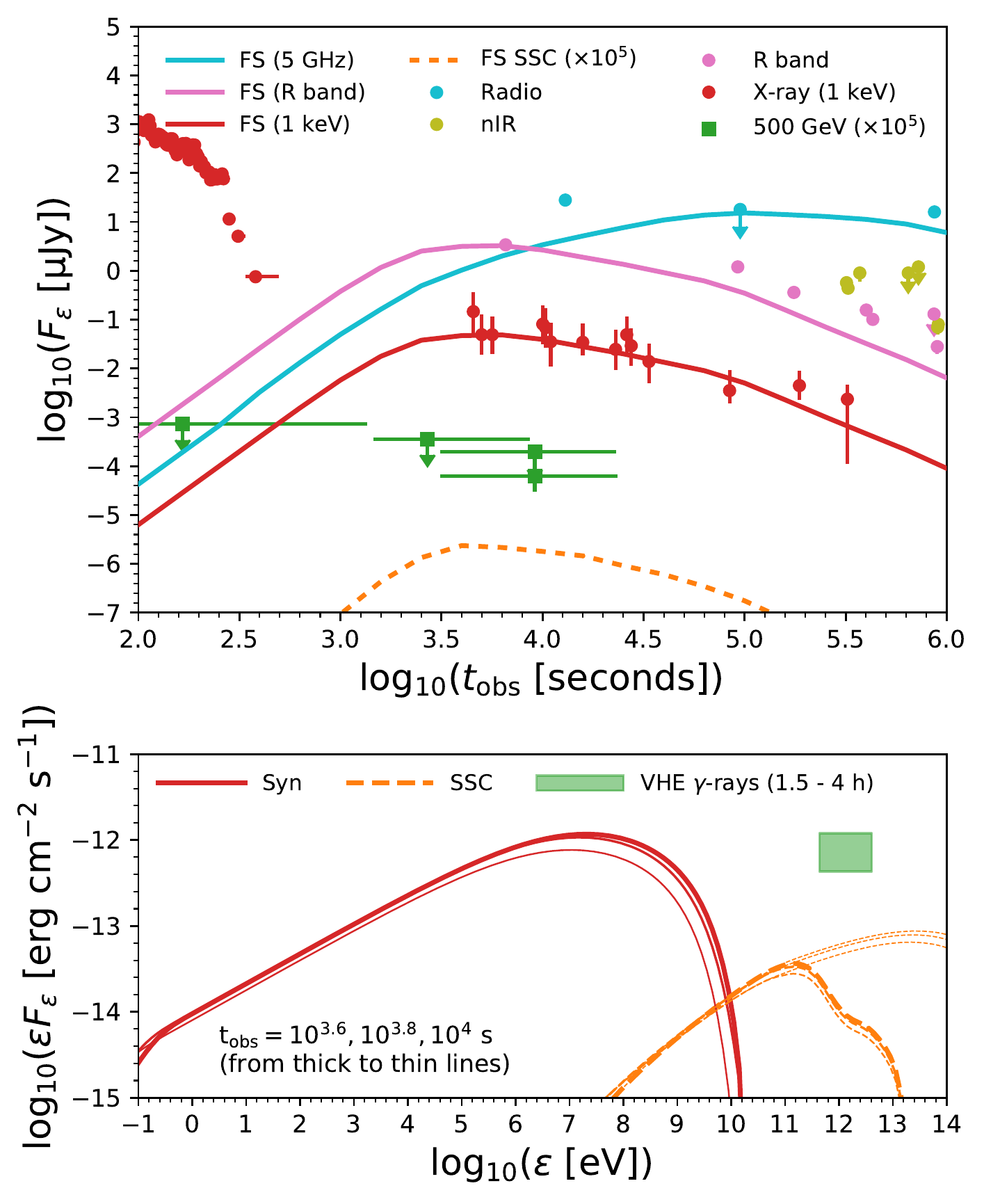}
\caption{Upper panel: Multi-wavelength light curves from radio, optical, X-ray to VHE bands at 500 GeV in the SSC scenario. The radio and optical are taken from~\cite{Troja:2019ccb,Lamb:2019lao}, late-time X-ray data are taken from~\citep{Acciari:2020ljs}, while the early X-ray data at extended emission phase are taken from the public on-line repository assuming spectral index $\Gamma_X = 2.5$~\citep{Evans2010}.
The green dots are the upper limit of the flux at 500 GeV assuming a power-law spectrum with spectral index $\Gamma_\gamma = 2$~\citep{Acciari:2020ljs}.
Lower panel: Energy spectra at $t = 10^{3.6}-10^4~\rm~s$ for different components, synchrotron emission and SSC emission. The thick dashed lines take into account both of the internal $\gamma \gamma$ absorption and EBL attenuation~\citep{Kneiske:2003tx}, while the thin lines ignore the EBL attenuation. The rectangle band show the integrated flux level measured by MAGIC from 1.5 hour to 4 hour~\citep{Acciari:2020ljs}. The physical parameters are $\mathcal{E}_k = 3 \times 10^{51}\rm~erg$, $\Gamma_0 = 40$, $n_{\rm ex} = 0.05\rm~cm^{-3}$, $\epsilon_e = 0.3$, $\epsilon_B = 10^{-5}$, $f_e = 0.5$, $s = 2.3$, and $\theta_j = 0.15$.
\label{fig:GRB160821B_ModelSSC_1_lightcurve}}
\end{figure}

In Fig.~\ref{fig:GRB160821B_ModelSSC_1_lightcurve}, we show the light curves (upper panel) and energy spectrum (lower panel) of the afterglow emission predicted in the SSC scenario.
In the upper panel, we show multi-wavelength light curves from the external forward shock. 
The early radio data can be explained by the emission from reverse shock and the late optical to infrared data may have contributions from a kilonova ejecta~\citep{Lamb:2019lao, Troja:2019ccb}.
In the lower panel, we show broadband energy spectra for synchrotron and SSC components at $t \sim 10^{3.4} - 10^4\rm~s$ .
The thick $\gamma$-ray energy spectra take into account the effect of $\gamma \gamma$ absorption, where the extragalactic background light (EBL) attenuation is more important than the internal absorption by ambient photons.
We also added the MAGIC light curve and the energy spectrum with the EBL correction for comparison~\citep{Acciari:2020ljs, Inoue:2019ICRC}.

We see, however, that from the SSC mechanism the predicted flux of \vhe is $\sim 1-2$ orders of magnitude lower than the MAGIC detection threshold within the time range from $t \sim 10^3\rm~s$ to $t \sim 10^4\rm~s$. The Compton Y parameter, which is expressed as the ratio of the SSC power to the synchrotron power, can be estimated to be $Y_{\rm SSC}(\gamma_e) \propto U_{\rm syn}^\prime[\varepsilon_\gamma^\prime < \varepsilon_{\rm KN}^\prime] / U_B^\prime$, where primed quantities are measured in the jet comoving frame, $U_{\rm syn}^\prime$ is the comoving synchrotron photon density, $\varepsilon_{\rm KN}^\prime$ is the comoving Klein-Nishina break energy~\citep{Murase:2010fq}, and $U_B^\prime = B^2 / 8\pi$ is the comoving magnetic energy.
We find that the value of $Y_{\rm SSC}$ is lower than what is expected in the Thompson regime where $Y_{\rm SSC} \sim \sqrt{(\epsilon_e / \epsilon_B)}$~\citep{Sari:2000zp}, which means that the SSC cooling is suppressed in the Klein-Nishina regime when the photon energy measured in the electron rest frame is comparable to the electron rest mass.
Thus, the SSC process is suppressed and enters into the deep KN regime given the parameters used in Fig.~\ref{fig:GRB160821B_ModelSSC_1_lightcurve}. 
We conclude that it is challenging for the SSC scenario to achieve an energy flux level suggested by the MAGIC data.      
However, the situation is different if we consider the EIC scenario, where the external photons, e.g., coming from late-prompt emission, are more copious than synchrotron photons.

\section{\vhe in the external inverse-Compton scenario}\label{sec:EIC-scenario}
In this section, we present the EIC scenario, in which seed photons are attributed to extended and plateau emission.
We fit the observed light curve of \grb using a phenomenological formula for extended emission~\citep{Kisaka:2017tas},
\begin{equation}\label{extended}
    L_{\rm EE}(t) = L_{b, \rm EE} \left(1 + \frac{t}{t_{\rm EE}}\right)^{-\delta_{\rm EE}}
\end{equation}
where $L_{b, \rm EE} \simeq 6 \times 10^{48}\rm~erg~s^{-1}$ is the luminosity of the extended emission at the break energy $\varepsilon_{b, \rm EE}$, $t_{\rm EE} \simeq 4 \times 10^2\rm~s$ is the duration of the extended emission, and $\delta_{\rm EE} \simeq 10$ is the decay index which reflects the sharp decline of the extended emission~\citep{Troja:2019ccb}.
For plateau emission,
\begin{equation}\label{plateau}
    L_{\rm PL}(t) = L_{b, \rm PL} \left(\frac{t}{t_{\rm PL}}\right)^{-\gamma_{\rm PL}} \left(1 + \frac{t}{t_{\rm PL}}\right)^{-\delta_{\rm PL}},
\end{equation}
where $L_{b, \rm PL} \simeq 4 \times 10^{43}\rm~erg~s^{-1}$ is the luminosity of the plateau emission at peak energy $\varepsilon_{b, \rm PL}$ , $t_{\rm PL}\simeq 2 \times 10^5\rm~s$ is the duration of the plateau emission, $\delta_{\rm PL} = 20/3$ is the decay index to describe the decline of the plateau emission, and $\gamma_{\rm PL} = 1/2$ is a factor to describe the gradual decline of the plateau emission.
The index $\delta_{\rm EE/PL}$ depends on the temporal evolution of the central engine activity. For example, the value of $\delta_{\rm EE/PL} \sim 40/9$ can be derived assuming the fall-back accretion rate, $\dot{M} \propto t^{-5/3}$~\citep{Kisaka:2015sya}.
We find that the X-ray light curve shows a sharp decline at the end of the extended and plateau emission, and $\delta_{\rm EE/PL} > 40/9$ gives a better fit to the observation data, which could be realized by e.g., the decay of magnetic fields via magnetic reconnections~\citep{Kisaka:2015sya}. 
The energy spectrum of late-prompt extended and plateau emission can generally be described by a broken power law $G(\varepsilon) \propto{(\varepsilon / \varepsilon_b)}^{-\alpha + 1}$ for $\varepsilon < \varepsilon_b$ and $G(\varepsilon) \propto {(\varepsilon / \varepsilon_b)}^{-\beta + 1}$ for $\varepsilon > \varepsilon_b$, where $\varepsilon_b$ is the break energy, $\alpha$ and $\beta$ are the spectral indices that depend on the details of the emission mechanism~\citep{Kumar:2014upa}, such as the dissipative photosphere~\citep{Rees:2004gt}, internal shock~\citep{Rees:1994nw}, or magnetic dissipation~\citep{Zhang:2010jt}. 
For simplicity, we assume that the late-prompt emission has an energy spectrum with a peak energy of $\varepsilon_{b, \rm EE} = 4\times 10^3 \rm~eV$, $\alpha_{\rm EE} = 0.5$, $\beta_{\rm EE} = 2.6$, $\varepsilon_{b, \rm PL} = 30 \rm~eV$, $\alpha_{\rm PL} = 0.2$ and $\beta_{\rm PL} = 2.7$ ~\citep{Kagawa:2019hvt}.
Although the X-ray flux may overshoot the data around $10^{2.5}$~s, the results are not much affected because seed photons are mainly supplied by plateau emission thanks to the Klein-Nishina effect.   

\begin{figure}
\includegraphics[width=\linewidth]{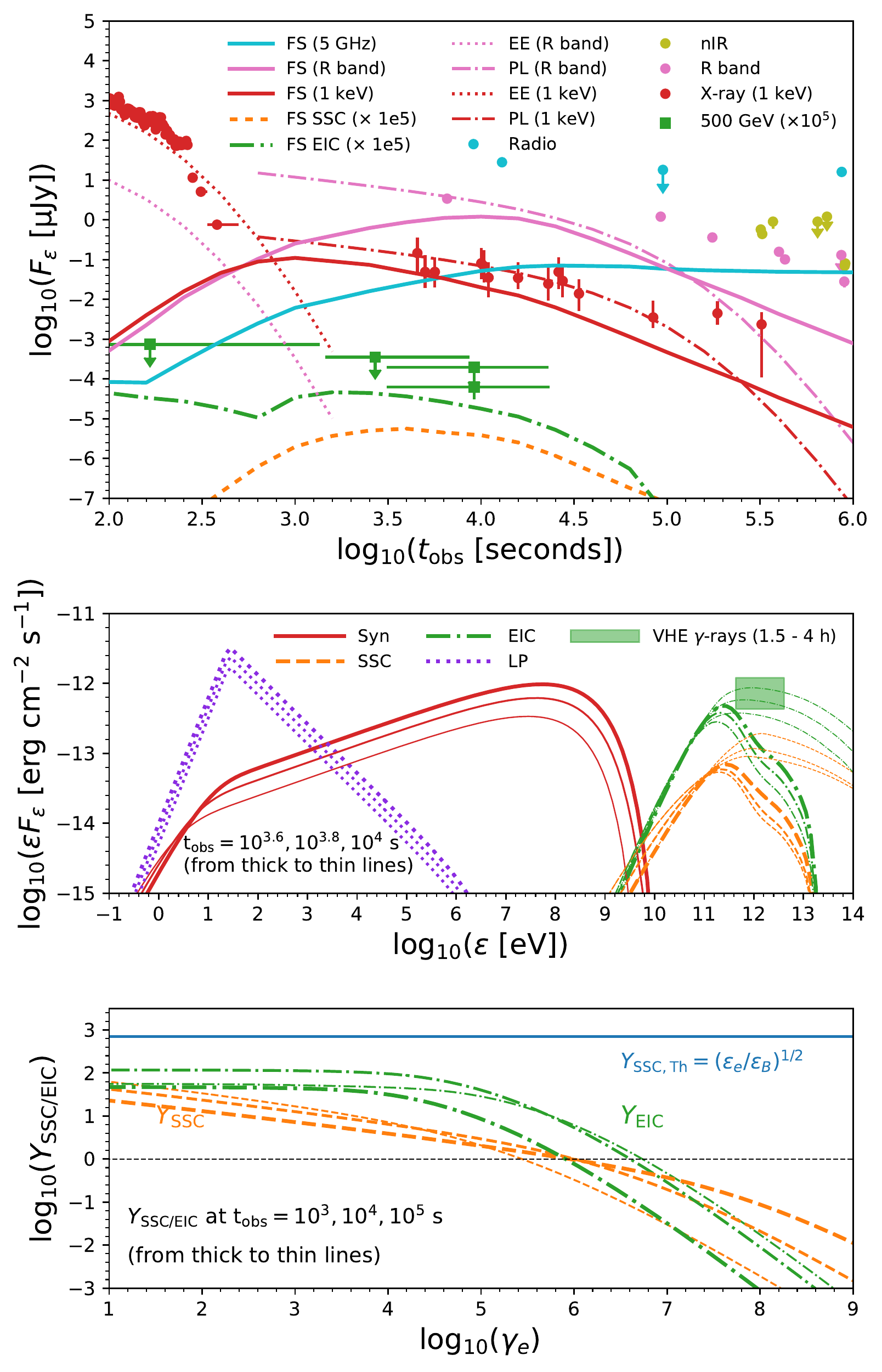}
\caption{Top panel: multi-wavelength light curves from radio, optical, X-ray to VHE band at 500 GeV in the EIC scenario. The data shown are the same as in Fig.~\ref{fig:GRB160821B_ModelSSC_1_lightcurve}. 
We include late-prompt emission consisting of extended and plateau emission fitted with Eq.~\ref{extended} and Eq.~\ref{plateau}, and \vhe from the EIC process. Middle panel: same as Fig.~\ref{fig:GRB160821B_ModelSSC_1_lightcurve}, we show the broadband spectra for $t_{\rm obs}=10^{3.6}-10^4$ s. Bottom panel: Compoton Y parameter, $Y_{\rm SSC}$ (dashed line) and $Y_{\rm EIC}$ (dash-dotted line), as a function of electron Lorentz factor $\gamma_e$, and the blue solid line is the $Y_{\rm SSC}$ estimated in the Thompson regime. The physical parameters are $\mathcal{E}_k = 2 \times 10^{51}\rm~erg$, $\Gamma_0 = 60$, $n_{\rm ex} = 0.1\rm~cm^{-3}$, $\epsilon_e = 0.5$, $\epsilon_B = 10^{-6}$, $f_e = 0.1$, $s = 2.5$, and $\theta_j = 0.1$. 
\label{fig:GRB160821B_Model}}
\end{figure}

As shown in Fig.~\ref{fig:GRB160821B_Model}, our EIC scenario can explain the $\sim 3 \sigma$ detection by the MAGIC telescopes, considering observation uncertainties and poor weather conditions at earlier times. One can see that the SSC light curve is negligible compared to the EIC light curve (dash-dotted line).
The EIC light curve reaches a peak at the end of the extended emission because of the significant deceleration of the outflow at the deceleration time $t_{\rm dec}$, where more non-thermal electrons are accumulated and IC scatterings cause time delays compared to seed photons impinging behind.
Even though the EIC light curve is usually flatter than the SSC light curve~\citep{Murase:2009su,Murase:2010fq}, the transition of the seed photons coming from the extended emission to those from the plateau
emission is seen in the EIC light curve.
The predicted flux of \vhe can reach $\sim 10^{-12}\rm~erg~cm^{-2}~s^{-1}$ at $10^3 - 10^4\rm~s$.
In Fig.~\ref{fig:GRB160821B_Model}, we also show the corresponding Compton Y parameters, $Y_{\rm SSC}(\gamma_e)$ and $Y_{\rm EIC}(\gamma_e)$ as a function of electron energy.
We find that the IC scattering with late-prompt plateau photons are between the Thompson regime and KN regime for electrons with Lorentz factors of $\gamma_e \sim 10^4 - 10^6$. This is because the photon energy density $U_{\rm ph}^\prime[\varepsilon_\gamma^\prime < \varepsilon_{\rm KN}^\prime]$ is insensitive to $\varepsilon_{\rm KN}^\prime$.
Thus, we can expect EIC emission brighter than SSC emission.

The light curves in optical and near-infrared bands of \grb have a bump at $\sim 1$ day, which is interpreted as a kilonova/macronova component with a temperature of $T_{\rm KN,day}\approx4500$ K~\citep{Troja:2019ccb,Lamb:2019lao}.
If the plateau emission is the internal origin and its dissipation region is inside the kilonova ejecta, electrons accelerated in the prolonged relativistic jets can emit high-energy gamma-rays by upscattering of kilonova photons~\citep{Kimura:2019fae}. The thermal photon energy in the kilonova can be approximated to be $T_{\rm KN}(t)\approx T_{\rm KN,day}(t/\rm day)^{-0.8}\simeq2.8\times10^4$~K at the time of MAGIC detection. Then, the KN break energy of the upscattered kilonova photon spectrum is estimated to be $\varepsilon_{\rm KN}\approx m_e^2c^4/(2.8k_BT_{\rm KN})\sim40$ GeV. This is well below the MAGIC threshold energy, and it is thus challenging to explain MAGIC data by the upscattered kilonova photons.

\section{Summary and Discussion}\label{sec:summary}
We studied the origin of \vhe from GRB 160821B. 
While the SSC scenario has difficulty in explaining the MAGIC data, our results showed that the predicted EIC flux can dominate over the SSC flux in the case of GRB 160821B. In the EIC scenario, the extended and the plateau emission is attributed to the late-prompt emission from the long-lasting central engine, which can provide seed photons necessary for the EIC process.
The resulting very-high-energy $\gamma$-ray flux can reach $\sim 10^{-12}\rm~erg~cm^{-2}~s^{-1}$, which is consistent with the MAGIC observation.

In general, the detection of \vhe from short GRBs is more challenging, because they are less energetic than long GRBs. This is especially the case in the SSC scenario. 
This work demonstrated that the very-high-energy signal can be enhanced by extended and plateau emission that can provide seed photons allowing the EIC emission to be dominant. 
One prominent feature of the EIC light curve that is accompanied with the extended and plateau emission is that the TeV peak is reached around the end of the seed photons. Such EIC \vhe are promising targets for future IACTs observing short GRBs and their off-axis emission are promising EM counterparts of GWs from NS mergers~\citep{Murase:2017snw}.
The detection of \vhe from short GRBs will also enable us to not only probe the activities of a central engine, which can be a black hole or NS, but also to constrain the environments of short GRBs and their host galaxies, and intergalatic magnetic fields in cosmic voids~\citep{Murase:2008yy}. 

\begin{acknowledgements}
We thank Ke Fang, Susumu Inoue, and Koji Noda for useful discussion. The work of K.M. is supported by the Alfred P. Sloan Foundation, NSF Grant No.~AST-1908689, and KAKENHI No.~20H01901 and No.~20H05852. B.T.Z. acknowledges the IGC fellowship. C.C.Y. and P.M. acknowledge support from the Eberly Foundation. S.S.K. acknowledges the JSPS Research Fellowship, JSPS KAKENHI Grant No.~19J00198. The early part of this work is supported by the Fermi GI program 111180. 
\end{acknowledgements}

\bibliographystyle{aasjournal}
\bibliography{bzhang}

\clearpage

\end{document}